# Run Control and Monitor System for the CMS Experiment

V. Brigljevic, G. Bruno, E. Cano, S. Cittolin, A. Csilling, D. Gigi, F. Glege, R. Gomez-Reino,
M. Gulmini[1,*], J. Gutleber, C. Jacobs, M. Kozlovszky, H. Larsen, I. Magrans, F. Meijers, E. Meschi,
S. Murray, A. Oh, L. Orsini, L. Pollet, A. Racz, D. Samyn, P. Scharff-Hansen, C. Schwick,
P. Sphicas[2]
*CERN, European Organization for Nuclear Research, Geneva, Switzerland*
*[1] Also at INFN, Laboratori Nazionali di Legnaro, Legnaro, Italy*
*[2] Also at University of Athens, Greece*

L. Berti, G. Maron, G. Rorato, N. Toniolo, L. Zangrando
*INFN, Laboratori Nazionali di Legnaro, Legnaro, Italy*

M. Bellato, S. Ventura
*INFN, Sezione di Padova, Padova, Italy*

S. Erhan
*University of California, Los Angeles, California, USA*

*presented by Michele Gulmini (Michele.Gulmini@cern.ch)

The Run Control and Monitor System (RCMS) of the CMS experiment is the set of hardware and software components responsible for controlling and monitoring the experiment during data-taking. It provides users with a "virtual counting room", enabling them to operate the experiment and to monitor detector status and data quality from any point in the world. This paper describes the architecture of the RCMS with particular emphasis on its scalability through a distributed collection of nodes arranged in a tree-based hierarchy. The current implementation of the architecture in a prototype RCMS used in test beam setups, detector validations and DAQ demonstrators is documented. A discussion of the key technologies used, including Web Services, and the results of tests performed with a 128-node system are presented.

## 1. INTRODUCTION

Modern Data Acquisition Systems (DAQs) are composed of several physically distributed cooperating devices that need to be configured, controlled and monitored during data-taking. A Run Control and Monitor System (RCMS) provides a set of user-friendly interfaces that allow the user to easily operate the experiment, hiding the complexity of the DAQ system.

The RCMS of the CMS experiment is defined as the collection of hardware and software components responsible for controlling and monitoring the CMS experiment during data-taking. In effect, the RCMS provides users with a "virtual counting room", enabling them to operate the experiment and monitor detector, DAQ status and data quality from anywhere in the world.

The architecture of the CMS DAQ system [1] assumes that there are roughly O(10^4) objects that need to be controlled. The RCMS architecture must clearly be scalable to effectively control and monitor a large distributed system.

Current Internet applications face similar issues as the RCMS. Heterogeneous distributed systems need to be interconnected. Many clients distributed in a large geographical area must interoperate with a set of servers and databases. Scaling up in terms of database transactions per second and connectivity to the services, providing at the same time secure access, are main requirements. Most Internet applications have adopted Web technologies and in particular Web

Services [2]. The XML data format [3] is commonly used for data exchange, and the SOAP protocol [4] for communication.

Since the main requirements of the CMS RCMS are similar to those of Internet applications, the CMS system has adopted the same technologies. The use of standard or otherwise widely adopted technologies allows for maximum profit of advances made in the Internet world.

The following section presents the architecture and design of the RCMS, followed by a general overview of the software technologies used in a number of prototypes and evaluated for adoption in the final system. Finally a summary of the current status of the development is presented.

## 2. RCMS

As a component of the CMS online software, the RCMS is designed to interoperate with the other online software components, like the Detector Control System (DCS), and the XDAQ cross-platform DAQ framework [5,6].

The RCMS views the experiment as a collection of partitions. A partition is a configurable group of resources. Multiple partitions can be active concurrently, sharing resources if necessary, allowing several sections of the experiment to run independently.

The RCMS performs actions on the partitions. Configuration and monitoring actions have timing constraints. Configuration and setup of partitions will require a time period of the order of minutes, their control (state change, execution of commands) of the order of seconds, while monitoring will range from microseconds to minutes depending on the amount of data required and their priority.

Users have to be authorized and authenticated to get access to the RCMS functions. Multiple users can access the experiment concurrently, performing subsets of possible functions exported by the system.

Hierarchies of distributed control applications will be used to address the $O(10^4)$ objects of the CMS DAQ system.

## 2.1. Architecture and Design

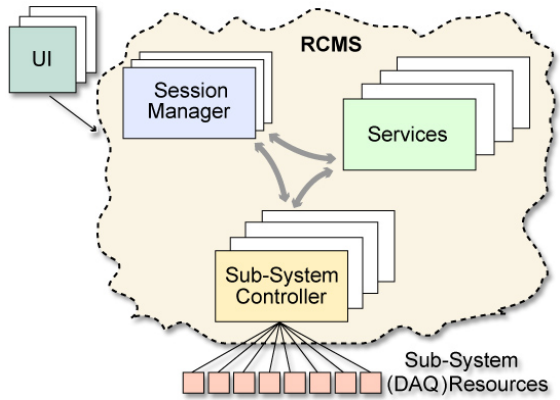

Figure 1: RCMS logical layout.

The logical layout of the RCMS is shown in figure 1. It consists of four types of elements: Session Managers, Sub-System Controllers, User Interfaces (UI) and a set of Services that are needed to support specific functions like security, logging and resource management.

A number of DAQ sub-systems are involved in data-taking (see figure 2). Any sub-system can be partitioned, and different partitions can be operated concurrently. Logical partitions managed by the RCMS have to satisfy a number of rules and constrains. The subdetectors, e.g. like the muon detectors, the tracker and the calorimeters, represent "natural" partition boundaries. DAQ sub-systems, like parts of the Event Builder and the Detector Control System (DCS), which deal directly with the detectors and their electronics have to be partitioned according to the detector they are attached to.

The DCS sub-system, from the point of view of the RCMS, is an external system with its own independent partition management. During data taking, the RCMS takes control, instructing the DCS to set up and monitor the partitions corresponding to the detector elements involved in the acquisition session.

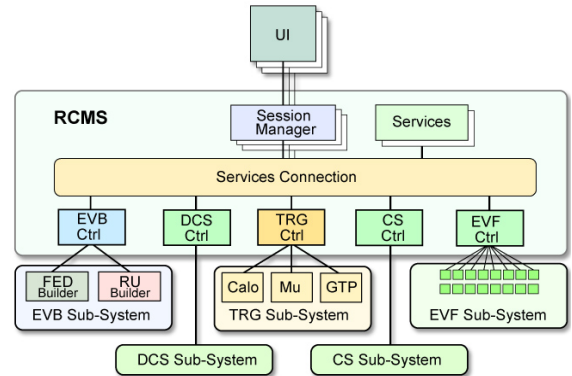

Figure 2: Session Managers and Sub-Systems defined in the RCMS.

The DAQ sub-systems shown in figure 2 can be further decomposed into other sub-systems. Sub-system partitions are organized into a hierarchy to form larger partitions. Actions performed on these larger partitions are propagated to their component partitions.

The running of a partition is defined as a "session". Multiple sessions may coexist and be active concurrently. Each session is associated with a Session Manager (SMR) that coordinates user access to the session and propagates commands from the users to the Sub-System Controllers (SSCs). High-level commands are interpreted and expanded into sequences of commands for the proper sub-system resources.

The SSC consists of a number of Function Managers (FMs) and local database services. One Function Manager is instantiated for each partition. A FM receives requests from the corresponding Session Manager and transforms them into commands that are sent to the sub-system. A local database service can be used as proxy/cache to the other RCMS services to facilitate software and configuration download, monitoring and logging operations.

The results of performed actions are logged and analyzed by the RCMS. Information concerning the internal status, monitor data and error messages from the various DAQ subsystems is also handled by the RCMS.

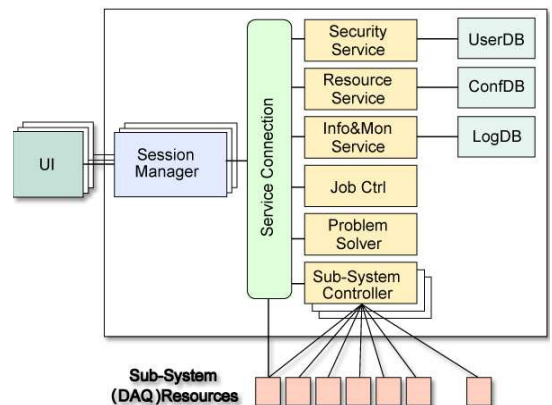

Figure 3: Block Diagram of the Run Control and Monitor System.

A number of services support the interaction with users to manage sub-system resources.

Figure 3 shows a block diagram of the RCMS with the services defined so far.

The Security Service provides facilities for user authentication and authorization, including data encryption when demanded. It manages the user database where profiles are stored, including access rights and a working history.

The Resource Service (RS) (see figure 4) handles all the resources of the DAQ, including partitions. A resource can be any hardware or software component involved in a DAQ session. Resources can be discovered, allocated and queried. Partitions can only use available resources. It is the responsibility of the RS to check resource availability and contention with other active partitions when a resource is allocated for use in a DAQ session. A periodic scan of the registered resources will keep the configuration database up to date.

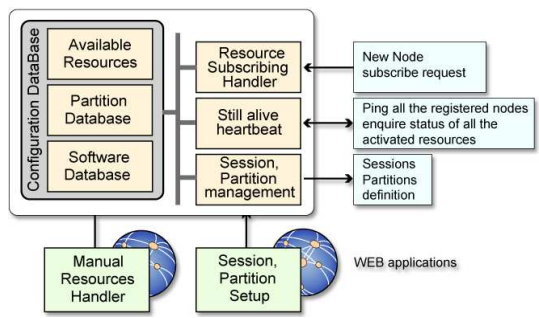

Figure 4: Block diagram of the Resource Service.

The Information and Monitor Service (IMS) (see figure 5) collects messages and monitor data coming from DAQ resources and RCMS components and stores them in a database. There are several types of messages collected from the sub-systems. The messages are catalogued according to their type, severity level and timestamp. Data can be provided in numeric formats, histograms, tables and other forms. The IMS collects and organizes the incoming information in a database and publishes it to subscribers. These subscribers can register for specific messages categorized by a number of selection criteria, such as timestamp, information source and severity level.

The Job Control (JC) is a utility that allows the remote execution and supervision of any software process involved in data-taking. It is used to start, stop and monitor the software infrastructure of the RCMS and the DAQ software components.

The Problem Solver (PS) identifies malfunctions of the DAQ system and determines possible recovery procedures. It subscribes to the IMS to receive the information it is interested in. The information is processed by a correlation engine and the result is used to determine a potential automatic recovery action, or to inform the system operator providing any analysis results it may have obtained.

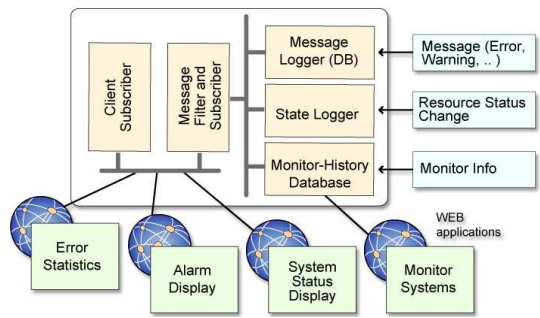

Figure 5: Block diagram of the Information and Monitor Service.

The above services can be distributed over a number of computing nodes for scalability and fault tolerance reasons. Each sub-system might have its own local instance of a service, thus resulting in a hierarchical architecture of services.

A number of Web applications are needed to operate the system. Graphical User Interfaces (GUIs) are very important to provide the user with a set of intuitive functions to control and monitor the experiment. All the services have their own administration and user GUIs. The Resource Service has to provide GUIs for storing new resources, defining and browsing partitions. The Information and Monitor Service must provide displays for alarm messages, error statistics, system status and monitoring.

## 2.2. Software Technologies

Web technologies and related developments play a fundamental role in the implementation of the proposed design. The XML data format and the SOAP communication protocol have already been mentioned in the introduction. The Web Services Description Language (WSDL) [7] is used to export service interfaces to clients. The Universal Description, Discovery and Integration (UDDI) registries [8] are used to locate service instances.

A rich choice of tools and solutions based on the previously mentioned Web technologies is available. Sun Java Web Services Developer Pack (JWSDP) [9], Apache Axis [10], and Glue [11] are widely used integrated development tools for Web Services. The Grid [12] community is considering the same technologies. The Globus [13] project has now adopted the Open Grid Services Architecture (OGSA) [14,15] and new releases of the toolkit are Web Services oriented.

Databases are largely used to store different types of information. Relational database management systems provide the required level of robustness and scalability. Native XML database technologies, based on the XMLDB interface [16] are an emerging, but not yet mature technology (see paragraph 4.3). They are taken into consideration because they remove the need to transform between XML data and relational table structures.

Expert systems technologies are being evaluated for the implementation of the Problem Solver service. Different types of input information need to be correlated in order to propose appropriate recovery actions. Commercial and public domain expert systems are available [17,18] and are currently being evaluated.

Security issues are crucial for a system such as the RCMS that has clients distributed around the globe. The RCMS is going to adopt one of the existing solutions available in the scope of large distributed systems. A candidate is the security infrastructure of the LHC Computing Grid (LCG) project [19].

## 2.3. Prototypes

A number of prototypes have been developed in order to test and validate the proposed architecture and its design. A fully functional prototype, called "RCMS for Small DAQ Systems", has been developed and used in several application scenarios of the CMS experiment. The goal is to provide a run control system for detector component testing and testbeam DAQs, and to verify the functionalities of the technologies chosen for its implementation. In the next section we present its features in more detail.

Other partially functional prototypes, called demonstrators, have been set up in order to investigate and validate the scalability of the system, to measure its performance and to evaluate the technologies used. In section 4 we describe the demonstrators that have been developed and the results obtained.

## 3. RCMS FOR SMALL DAQ SYSTEMS

The current prototype for small DAQ systems has been entirely developed using the Java programming language. It makes use of the Sun Java Web Services Developer Pack (JWSDP) [9] software tools. All the services are developed as Java servlets that run in a tomcat servlet container [20]. Presently the prototype includes the Resource Service, the Information and Monitor Service, the Session and Function Managers and a set of customizable GUIs. Support for both relational databases (mySQL) [21], and native XML databases (eXist) [22] has been provided. Figures 6 and 7 show the Resource Service prototype and the Information and Monitor Service prototype.

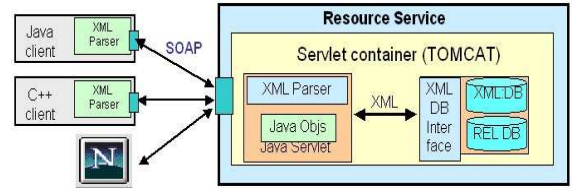

Figure 6: The Resource Service prototype.

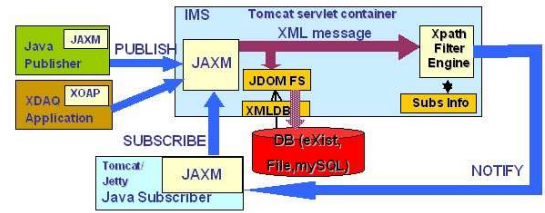

Figure 7: The Information and Monitor Service prototype.

The clients are Web applications able to communicate via SOAP over the http communication protocol. In particular they can be either XDAQ applications [23] or Web GUIs. Clients can publish messages to the IMS. Published messages are stored in a database and forwarded to a number of subscribers. Subscribers must be able to receive SOAP/http messages asynchronously. Both XDAQ applications and Java applications with a Jetty [24] embedded http server have been used.

The graphical user interface is based on Java applets and can run in current web browsers. It is composed of a generic framework, released together with the services, containing the basic functionalities common to any small CMS DAQ. These include the interface to the Resource Service for storing resources, partitions and DAQ configuration, the ability to subscribe to the IMS for visualizing errors and monitor information, the ability to command DAQ applications, and the support for the tcl scripting language [25]. To fit the specific requirements of each DAQ, the GUI functionality can be extended by dynamically loading new applets. Such applets must implement an API that has been developed in order to transparently integrate them in the GUI framework.

The "RCMS for Small DAQs" is currently used in some application scenarios of the CMS experiment [23]. Permanent installations are used for the test and validation of the CMS muon chambers [26] and for the tracker system test [27]. Temporary installations are used for testbeam DAQs, both for tracker and muon chambers, and DAQ demonstrators [28].

Figure 8 shows the GUI recently used at CERN for the muon chamber testbeam.

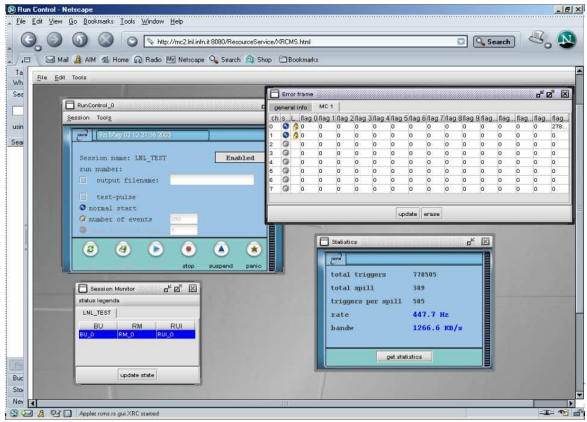

Figure 8: The Graphical User Interface for the Muon Chamber testbeam.

## 4. RCMS DEMONSTRATORS

Several demonstrators have been set up to prove the scalability of the RCMS architecture and to evaluate the technologies it uses. A Gigabit Ethernet based cluster of 128 dual-processor 1GHz Pentium-III 512 MB memory PCs has been used.

### 4.1. Hierarchy of Function Managers

A first demonstrator uses Function Managers developed for the RCMS for small DAQ systems in order to explore the ability to command a set of DAQ nodes running XDAQ applications.

Up to 120 XDAQ applications, each one running on a different node, perform a dummy action when receiving a SOAP control command, modifying their state and returning it to the Function Manager. Both flat and a two-level hierarchy of FMs have been investigated (see figure 9). Each FM is deployed on a different node of the cluster.

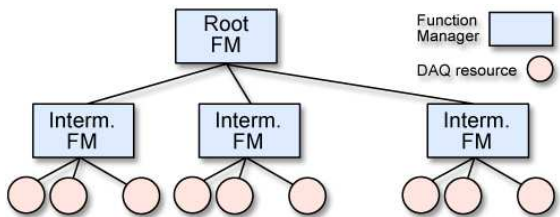

Figure 9: PC nodes organized in a hierarchical structure with an intermediate layer.

The time measured as a function of the number of nodes addressed represents the time required to perform a status change of the entire cluster.

Figure 10 shows three curves. The first one (A) is obtained when a single FM sends the commands sequentially to all the nodes, waiting for an acknowledgement for each command, before sending the following one. The second curve (B) is obtained using a multithreaded implementation of the FM. 8 threads have been used, so that multiple nodes can be commanded concurrently. The third curve (C) is obtained using a two-level hierarchy of multithreaded FMs.

The results show that changing the status of the entire cluster of 120 nodes can be performed in about 100ms by using a two-level hierarchy. The results help build confidence that the technology used can satisfy the requirement of performing a change of status and to execute commands within roughly one second.

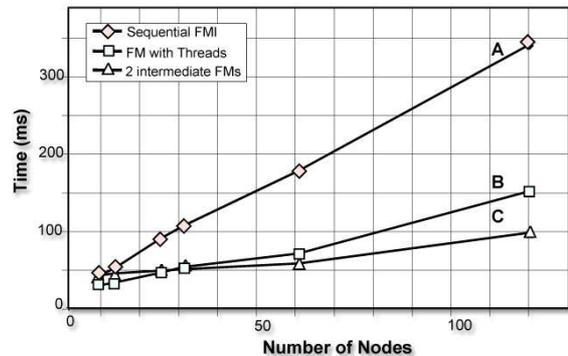

Figure 10: Time required to perform a state transition of a cluster of nodes running XDAQ applications.

### 4.2. Web Services

A second demonstrator has evaluated the Glue Web Services platform [11]. A simple logging service has been developed and several instances of the service have been deployed on physically distinct machines. Each instance exports its interface through the Web Services Description Language (WSDL) to a UDDI registry located in a dedicated host that publishes it to the clients. Clients looking for the logging service, query the UDDI registry to discover the location of the service and retrieve the access methods.

Rapid application deployment and run-time configuration and maintenance through the addition or removal of service instances via UDDI are some advantages of this technology.

Up to 4 service instances and a fixed number of 15 clients have been used. As expected when the load of the service instances is balanced the performance of the logging service scales linearly with the number of instances available (Figure 11).

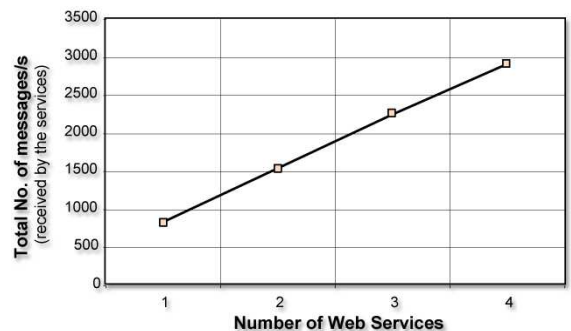

Figure 11: Performance of the log Web Service as a function of the number of service instances.

### 4.3.  Information and Monitor Service

A third demonstrator has been assembled using the same cluster of PCs and the current prototype implementation of the Information and Monitor Service used for small DAQs. The purpose of the first performance test is to understand the message rate the service can sustain. Up to 64 clients, called publishers, continuously send small (about 400 bytes) SOAP messages to the IMS.

The performance measured as a function of the number of messages managed by the IMS is shown in figure 12.

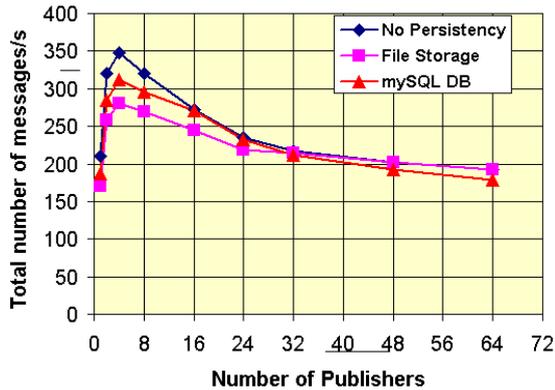

Figure 12: Total number of SOAP error messages managed by the IMS prototype as a function of the number of simultaneous publishers.

A single instance of the IMS prototype can manage between 200 and 300 SOAP messages per second, storing the error messages in local memory, on a flat file or using a mySQL database located on another PC. The mySQL database is accessed through JDBC (Java DataBase Connectivity API) [29].

The server reaches its saturation and best performance with 4 publishers. Using more publishers the performance decreases. Probably due to an overhead in the multithreading management of the tomcat servlet container this behaviour needs further investigation.

The performance using the eXist [22] XML database accessed through the XMLDB interface has also been measured. The result, not shown in the figure, is a few tens of messages per second, ten times worse than with a mySQL database or a flat file.

Another set of measurements has been made in order to verify the scalability of the prototype. Up to 4 instances of the IMS servlet and only one mySQL database accessed through JDBC have been used. The results are shown in figure 13. The total number of messages per second managed by the IMS service improves significantly with the number of instances of the IMS servlet. The maximum message rate is about 900 messages per second with four IMS instances. The performance is not limited by the mySQL database.

Figure 14 plots the total number of messages processed by the service when 16 publishers are used.

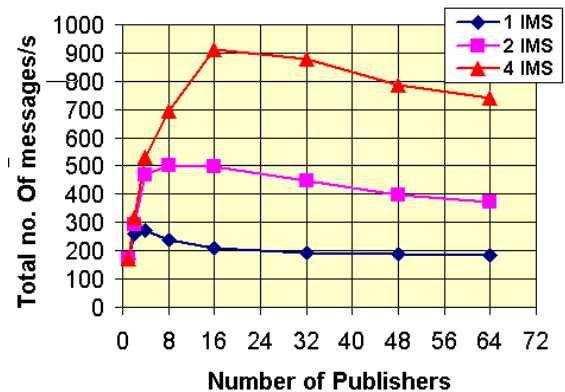

Figure 13: Total number of error messages managed by the IMS prototype depending on the number of service instances. A variable number of publishers is used.

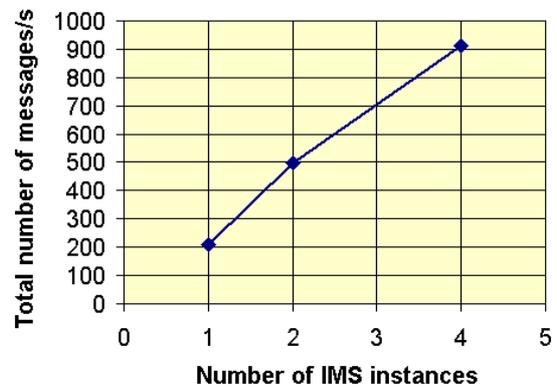

Figure 14: Total number of error messages managed by the IMS prototype as a function of the number of service instances. 16 publishers are used.

Preliminary measurements have also been made in order to understand the behaviour of the service when a number of clients subscribe for receiving information from the IMS. The preliminary test has shown a significant decrease in performance when adding subscribers to the system. Eight subscribers decrease the total message throughput by a factor of two. We are currently investigating the reasons of such behaviour. The code will be profiled in order to discover major inefficiencies, and a solution based on hierarchies of service instances has being designed.

### 5.  ONGOING WORK

The current RCMS for small DAQ systems is successfully in use in several CMS application scenarios. Experience obtained from the muon chambers and the tracker testbeam DAQs has suggested some improvements. Server side scripting language support, and graphical user interface improvements are some of them.

A Problem Solver prototype based on Jess (Java Expert System Shell) [18] is currently under development. This prototype is also being used to test the notification mechanism of the present IMS prototype.

To benefit maximally from the current trends in the Internet world, the Open Grid Services Architecture (OGSA) is under consideration. OGSA presents several commonalities with the RCMS. The Globus toolkit 3.0 (alpha release), which provides a first implementation of the architecture, is currently being evaluated. The Resource Service developed for small CMS DAQs has been ported to this framework. Functionality and performance tests are still in progress, with the main purpose of understanding if a smooth transition to the new architecture is possible.

Database management systems are a key component in the RCMS architecture. Both XML native databases and relational databases have been used in the prototypes. The prototypes did not attempt to test the scalability of the database technologies used. Further investigations and tests are foreseen.

## 6. SUMMARY

In this paper, the architecture and the design of the Run Control and Monitor System for the CMS experiment have been described. Facing similar requirements as Internet applications, Web Services technologies have been adopted. A control and monitoring system has been implemented for the small DAQ systems used for test beam setups, detector production centers and DAQ demonstrators. A number of RCMS demonstrators have also been developed in order to evaluate the Web technologies used as well as the scalability of the system. Preliminary results confirm the modularity and scalability of the design. Further investigations into Web Services technologies and database scalability are in progress. The goal of this work is a progressive smooth transition from the current software for small DAQs to the final implementation for the CMS experiment.